# First-Principles Investigation of Gas Adsorption on Bilayer Transition Metal Dichalcogenides for Sensing Toxic Gases

Jemal Yimer Damte[a], Hassan Ataalite[a]

[a] Department of Physics and NTIS - European Centre of Excellence, University of West Bohemia, Univerzitni 8 30614 Plzen Czech Republic

**Abstract**

Transition metal dichalcogenides (TMDs) have shown great promise in the field of gas sensing due to their high catalytic activity and unique electronic properties. They can effectively interact with various gas molecules, making them suitable materials for high-performance gas sensors. In this work, we have studied the sensing properties of nitrogen containing gases (NCGs) on different heterostructures (MoS/WTe, MoTe/WS, MoS/IrO, and MoS/TiO) using density functional studies. The result shows that $NH_3$ and NOx exhibit weak electronic interactions with MoS/WTe and MoTe/WS heterostructures and strong electronic interactions are observed between $NH_3$ and NOx molecules with MoS/IrO and MoS/TiO heterostructures. Electron transport properties were investigated using Non-Equilibrium Green's Function (NEGF) calculations. The result shows that MoS/WTe, MoTe/WS, MoS/TiO and MoS/IrO heterostructures exhibit promising characteristics as gas sensors for $NH_3$ detection. Additionally, MoS/WTe and MoTe/WS heterostructures show potential for detecting NOx. Based on this findings, it's clear that MoS/WTe and MoTe/WS heterostructures show great potential as gas sensors for both $NH_3$ and NOx detection, which has significant implications for various applications. This research provides valuable insights into the potential applications of different heterostructures for gas sensing, particularly in detecting nitrogen-containing gases and could pave the way for the development of highly sensitive and selective gas sensors with various practical applications.

**Introduction**

Nowadays, the world is grappling with air pollution and the greenhouse effect, leading to severe public health issues and environmental degradation [1-4]. The World Health Organization reports that millions of people die every year due to the presence of toxic gases in the air, causing damage to our circulatory and respiratory systems [5-8]. Nitrogen-containing gases such as nitrogen oxides (NOx) and ammonia ($NH_3$) are the primary pollutants in the environment, mainly emitted from the petrochemical industry and vehicles. Nitric oxide and nitrogen dioxide (NOx) are highly potent and have severe impacts on the metabolic activities of both humans and animals. NOx is a poisonous gas that can affect the human respiratory system and lead to various diseases even at low concentrations. Additionally, $NH_3$ reacts rapidly with other gases, posing a danger to public health by irritating the respiratory system, depositing in the lungs, and potentially causing loss of human life [9-13]. Therefore, detecting and capturing harmful gases is critically important for human well-being and the environment. This imperative has driven our efforts in designing new sensing materials to detect toxic gases. The presence of gases can be detected by a device called a gas sensor, which is essential for monitoring harmful toxic gases. Recently, there has been a surge in the design of massive gas sensor devices, playing a vital role in various applications including human health, environmental pollution monitoring, military and public safety, wearable devices, and smart farming [14-18]. To develop high-performance gas sensors, the sensing device should maintain high stability, sensitivity, and selectivity. As a result, researchers have investigated various sensing materials in recent studies to develop novel gas sensors. Among them, extensive research has been conducted on metal oxide gas sensors due to their inherent properties such as low cost, small particle size, and ease of production.

However, they have limited working conditions and stability, which hinders the gas sensor from achieving highly sensitive properties [19-25]. Therefore, it is essential to investigate gas sensing materials that operate at room temperature with high sensitivity, a large surface-to-volume ratio, and the ability to provide binding force for gas adsorption. In this regard, two-dimensional (2D) materials have been extensively studied owing to their large surface-to-volume ratio, higher mobility of charge carriers on the surface, and atomic-thin layered structures for high-performance sensing devices [20, 26-28]. Among different 2D materials, transition metal dichalcogenides (TMDs) exhibit excellent physical, chemical, and mechanical properties for nanoelectronic devices. Researchers have revealed that transition metal dichalcogenides are promising materials for sensing devices due to their unique properties such as easily tunable structures, numerous reactive sites, and high surface-to-volume ratio [29-33].

Both theoretical and experimental studies have reported that transition metal dichalcogenides are excellent toxic gas sensors [2, 34-36]. Gas sensing by transition metal dichalcogenides is based on the charge transfer process in the sensing mechanism. The latest research has focused on monolayer $MoS_2$- based sensors, which can detect $NO_2$ at sub-ppb levels (20 ppb) and $NH_3$ at 1 ppm [37-41]. Babar et al. have conducted investigations on monolayer and few-layer $MoS_2$ and $WS_2$ to understand their sensitivity and stable gas sensing characteristics when exposed to nitrogen- containing gases and carbon monoxide. Their report shows that bilayers and heterobilayers of $MoS_2$ and $WS_2$ have demonstrated excellent gas sensing performance [19, 21]. In our previous studies, we considered different heterostructures and designed a single device that incorporates a tribo-piezoelectric nanogenerator [42]. A tribo-piezoelectric nanogenerator is not only enhanced the power output but also ensured a stable power source. This is a valuable feature, especially for sensor applications where consistent power is crucial. In this study, we utilized selected heterostructures from our previous work to investigate toxic gas sensing characteristics. We have thoroughly examined the adsorption energies of all gases and electronic properties, such as electron density difference, Bader charge analysis, and density of states. Furthermore, to assess the sensitivity of gases to the selected structures, we studied the current-voltage characteristics before and after gas adsorption.

**Computational Details**

All density functional theory calculations were performed using the Spanish Initiative for Electronic Simulations with Thousands of Atoms (SIESTA) package [43]. As 2D heterostuctures are primarily held by dispersion forces, the van der Waals density functional(vdW-DF2) exchange functional has been considered into the calculations [44, 45]. We employed a double- zeta basis set with polarization functions (DZP), and electronic properties and geometry optimization were investigated using a Monkhorst–Pack grid method with a k-point sampling of 11x11x1. The kinetic energy mesh cutoff was set to 700 eV. Atomic positions were relaxed using a conjugate gradient (CG) algorithm until force and energy tolerances were below 0.01 eV / Å and $1 \times 10^{-5}$ eV, respectively. A vacuum-slab was utilized along the c-axis to prevent interactions among periodic images. The adsorption energy ($E_{ads}$) of gas molecules on TMD bilayers was calculated using the following formula:

$E_{ads} = E_{TMD/gas} - E_{TMD} - E_{gas.}$

Here, $E_{TMD/gas}$ represents the total energy of the system with gas adsorption, $E_{TMD}$ is the energy of TMD bilayers, and $E_{gas}$ is the energy of gas molecules. For investigating the electronic transport properties of the sensing systems, we calculated the current-voltage (I-V) characteristics using the TranSIESTA code, which employs the Non-Equilibrium Green's Function approach [46, 47]. Gold was used as the electrode material, with the scattering region and semi-infinite left and right electrodes forming part of the device. Both electrodes were connected to the scattering region. The transport electric current was calculated using the Landauer–Büttiker formula [48]. We have finally calculated the current flowing at constant bias condition using the post-processing tool TBTrans which is included in the TranSIESTA package.

**Results and discussion**

**Adsorption of NCGs on different heterostructures**

This work is a continuation of our previous research, where various homo and hetero bilayers were considered [42]. Among them, certain heterostructures were selected for tribo-piezoelectric nanogenerators. These selected heterostructures were then used for gas sensing applications. Nitrogen-containing gases were optimized on MoTe/WS, MoS/WTe, MoS/TiO, and MoS/IrO heterostructures, utilizing the favorable adsorption sites identified in previous studies [49, 50].

Table 1: Adsorption energy of NCGs, bond distance between gases and heterostructure and bader charge results.

| MoS/WTe | Eads (eV) | D (Å) | Q (e) |
|---|---|---|---|
| $NH_3$ | physisorbed | 3.04 | 0.02 |
| $NO_2$ | -1.02 | 2.80 | -0.49 |
| NO | -0.34 | 2.78 | -0.18 |
| MoTe/WS | Eads (eV) | D (Å) | Q (e) |
| $NH_3$ | physisorbed | 2.85 | 0.07 |
| $NO_2$ | -0.78 | 2.70 | -0.29 |
| NO | physisorbed | 2.60 | -0.02 |
| MoS/TiO | Eads (eV) | D (Å) | Q (e) |
| $NH_3$ | -0.68 | 2.41 | 0.19 |
| $NO_2$ | -1.67 | 2.33 | 0.01 |
| NO | -2.27 | 2.18 | 0.23 |
| MoS/IrO | Eads (eV) | D (Å) | Q (e) |
| $NH_3$ | -0.54 | 3.11 | 0.26 |
| $NO_2$ | -1.13 | 2.13 | 0.02 |
| NO | -1.92 | 1.87 | 0.24 |

The optimized structures of nitrogen-containing gases are shown in Figure 1-4, and the adsorption energies, along with geometric bond distances, are presented in Table 1. For $NH_3$, physisorption occurs in both MoTe/WS and MoS/WTe heterostructures. However, the adsorption energy of $NH_3$ in MoS/TiO and MoS/IrO heterostructures is -0.68 eV and -0.54 eV, respectively. The calculated adsorption distances between $NH_3$ and MoTe/WS, MoS/WTe, MoS/TiO, and MoS/IrO heterostructures are 2.85 Å, 3.04 Å, 2.41 Å, and 3.11 Å, respectively. In addition, the adsorption energy of $NO_2$ was calculated. $NO_2$ adsorption on MoS/TiO heterostructures has a larger adsorption energy of -1.67 eV. The adsorption energy of $NO_2$ on MoTe/WS, MoS/WTe, and MoS/IrO heterostructures are -0.78 eV, -1.02 eV, and -1.13 eV, respectively. The calculated adsorption distances between $NO_2$ and the heterostructures are 2.80 Å, 2.70 Å, 2.33 Å, and 2.13 Å in MoS/WTe, MoTe/WS, MoS/TiO, and MoS/IrO heterostructures, respectively. Regarding NO adsorption, it is physisorbed on MoTe/WS heterostructure but chemisorbed in MoS/WTe, MoS/TiO, and MoS/IrO heterostructures. The adsorption energies are -0.34 eV, -2.27 eV, and -1.92 eV in MoS/WTe, MoS/TiO, and MoS/IrO heterostructures, respectively. The calculated adsorption distances between NO and MoTe/WS, MoS/WTe, MoS/TiO, and MoS/IrO heterostructures are detailed in Table 1.

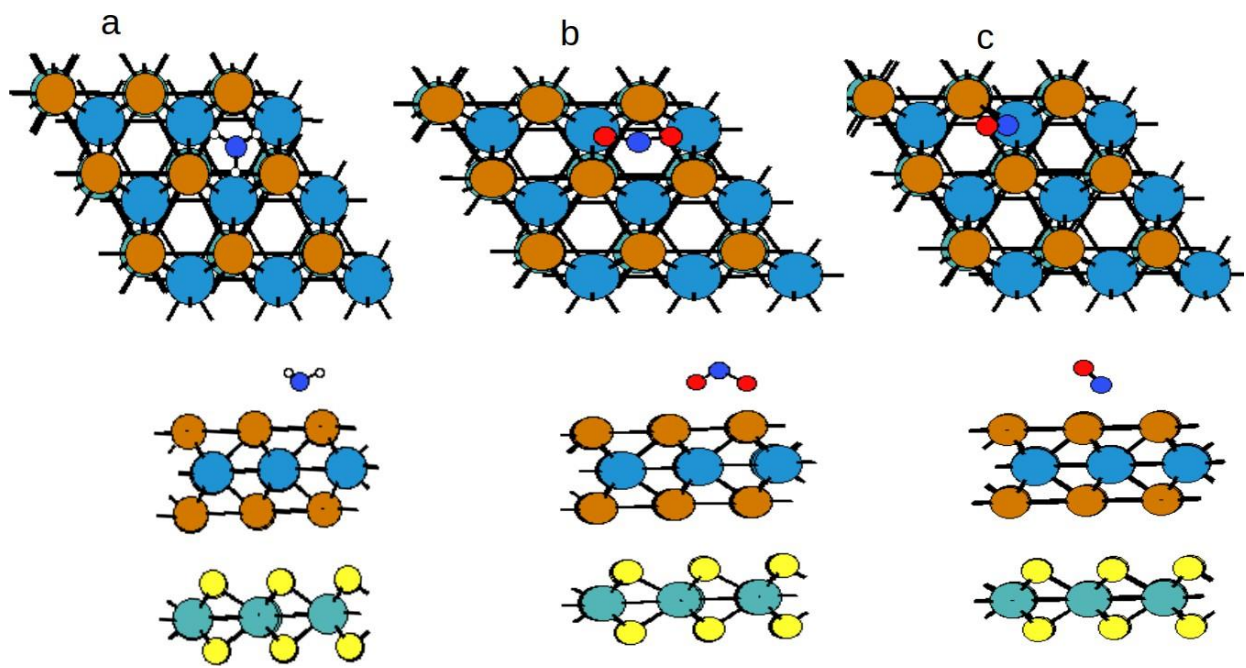

Figure 1: Optimized adsorption structures of a) $NH_3$, b) $NO_2$ and c) NO on MoS/WTe heterostructure.

The shorter distances between NO and MoS/TiO, and MoS/IrO heterostructures indicate stronger interactions. In summary, the adsorption energy of nitrogen-containing gases on MoS/TiO and MoS/IrO heterostructures is higher compared to the adsorption of nitrogen-containing gases on MoS/WTe and MoTe/WS heterostructures.

**Electronic property analysis**

The electronic property analysis such as bader charge analysis, electron density difference (EDD) and projected density of states (PDOS) have been considered to investigate the sensing mechanism of NCGs on MoS/WTe, MoTe/WS, MoS/TiO and MoS/IrO heterostructures.

**Bader charge analysis**

The Bader analysis method was employed to obtain the charges of nitrogen containing gases adsorbed on MoS/WTe, MoTe/WS, MoS/TiO, and MoS/IrO heterostructures. Positive and negative values indicate electron donation and acceptance, respectively. The calculated Bader charge values of $NH_3$ in MoS/WTe, MoTe/WS, MoS/TiO, and MoS/IrO heterostructures are 0.02, 0.07, 0.19 and 0.26 electron, respectively. This confirms that $NH_3$ acts as an electron donor in all heterostructures, and the charge transfer to the sensing materials is consistent with the adsorption energy.

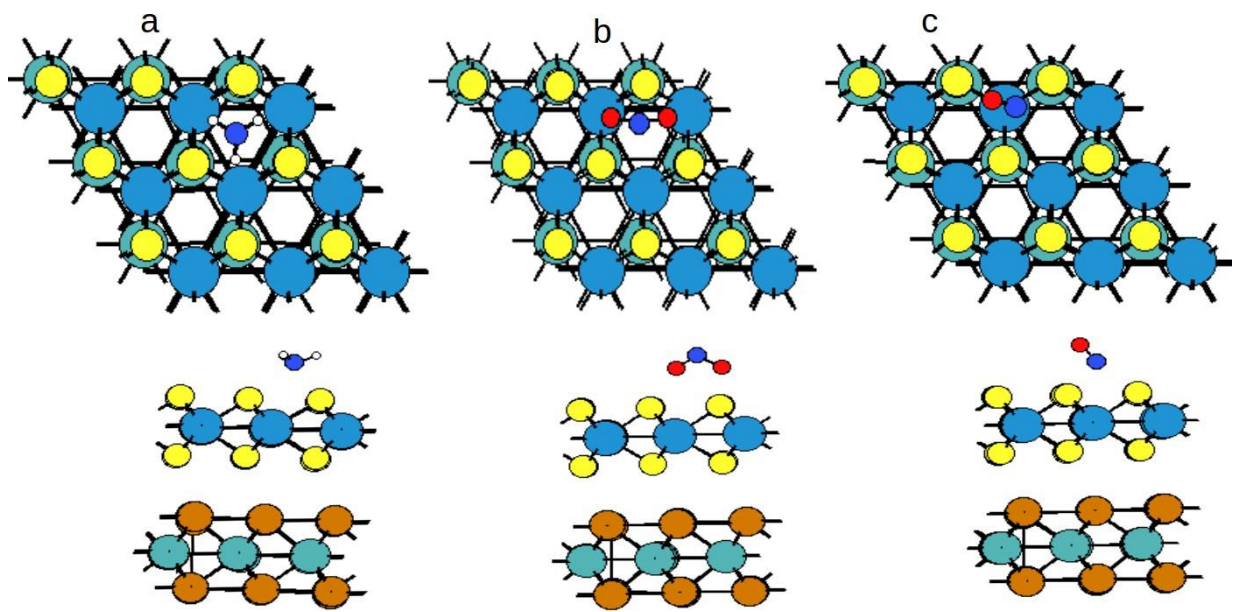

Figure 2: Optimized adsorption structures of a) $NH_3$, b) $NO_2$ and c) NO on MoTe/WS heterostructure

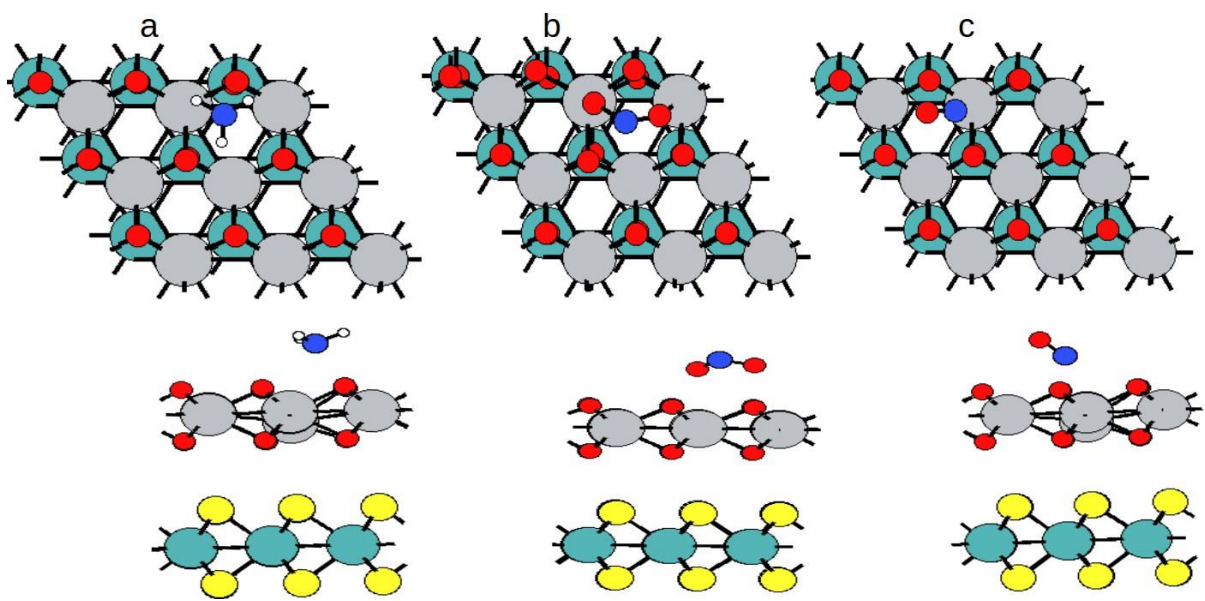

Figure 3: Optimized adsorption structures of a) NH$_3$, b) NO$_2$ and c) NO on MoS/TiO heterostructure

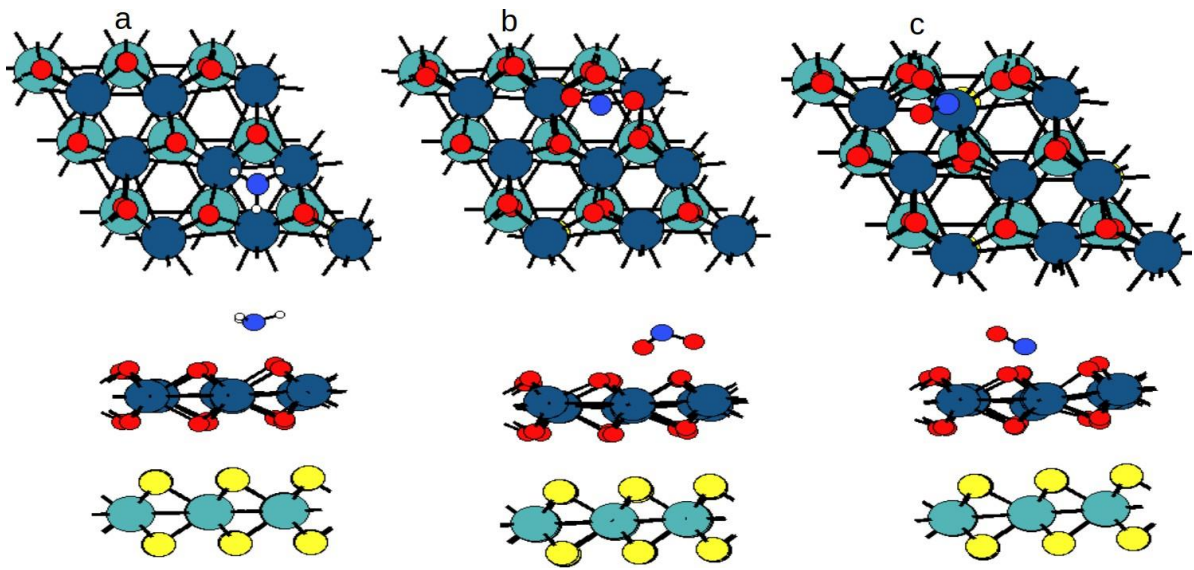

Figure 4: Optimized adsorption structures of a) NH$_3$, b) NO$_2$ and c) NO on MoS/IrO heterostructure

We observed that $NO_2$ and NO act as electron acceptors in MoS/WTe and MoTe/WS heterostructures, whereas they act as electron donors in MoS/TiO and MoS/IrO heterostructures. Furthermore, the results indicate that the adsorption energy of $NO_2$ in MoS/TiO and MoS/IrO heterostructures is not consistent with the charge transfer. Specifically, the charge transfer of $NO_2$ is very small compared to the charge transfer of NO in MoS/TiO and MoS/IrO heterostructures (see Table 1).

**Electron Density Difference (EDD) and Projected Density of States (PDOS) analysis**

We have further studied EDD and PDOS in order to study the orbital interactions of NCGs and the heterostructures. The electron density difference is calculated as follows:

$$\rho_{diff} = \rho_{total} - \rho_{heterostructure} - \rho_{molecule}$$

Where, $\rho_{total}$ is the electron density distribution of gas molecule adsorbed on the heterostructure, $\rho_{heterostructure}$ represents the electron density for the heterostructure and $\rho_{molecule}$ is the electron density of gas molecule. The EDD plots of nitrogen-containing gases on MoS/WTe, MoTe/WS, MoS/TiO, and MoS/IrO heterostructures are shown in Figure 5. Regions where charge depletion is observed are represented in cyan, while electron accumulation is indicated in yellow. Partial electron density depletion and accumulation have been observed around $NH_3$ in MoS/WTe, MoTe/WS, MoS/TiO, and MoS/IrO heterostructures. Additionally, the PDOS is plotted before and after the adsorption of $NH_3$ on the heterostructures. Small peaks appear in the conduction band after $NH_3$ adsorption, confirming electron transfer from $NH_3$ to the heterostructures. This result aligns with Bader charge calculations, where $NH_3$ acts as an electron donor in its adsorption on all heterostructures. Thicker electron density accumulation is observed.

after the adsorption of $NO_2$ on MoS/WTe and MoTe/WS heterostructures compared to its adsorption on MoS/TiO and MoS/IrO heterostructures. $NO_2$ accepts electrons from MoS/WTe and MoTe/WS heterostructures, consistent with Bader charge calculations where $NO_2$ acts as an electron acceptor in its adsorption on MoS/WTe and MoTe/WS heterostructures. Similar to $NH_3$, after the adsorption of $NO_2$, the peaks are shorter and overlap. Furthermore, the peaks shift towards the lower part of the valence band.

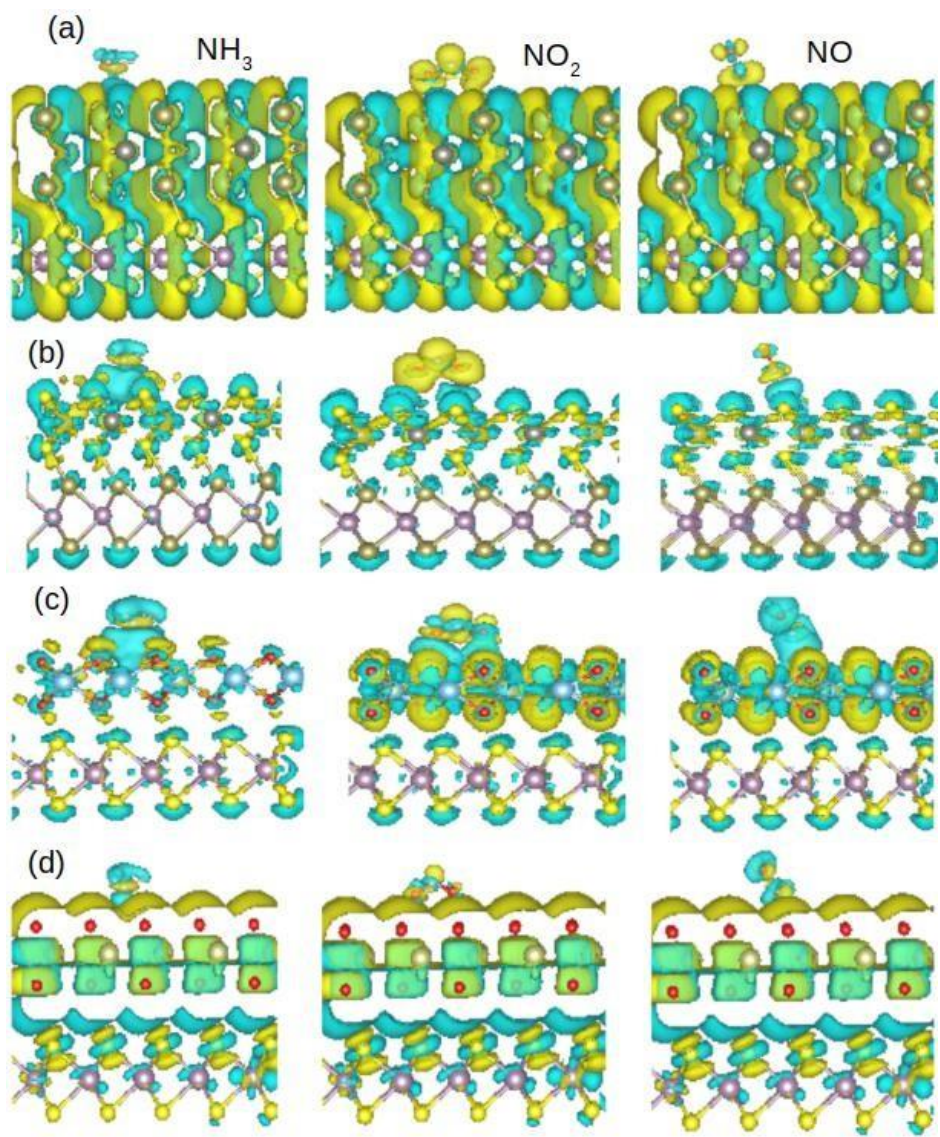

Figure 5: Calculated electron density difference on a) MoS/WTe, b) MoTe/WS, c) MoS/TiO

and d) MoS/IrO heterostructure. The iso surface level is 0.0005 e/ Å.

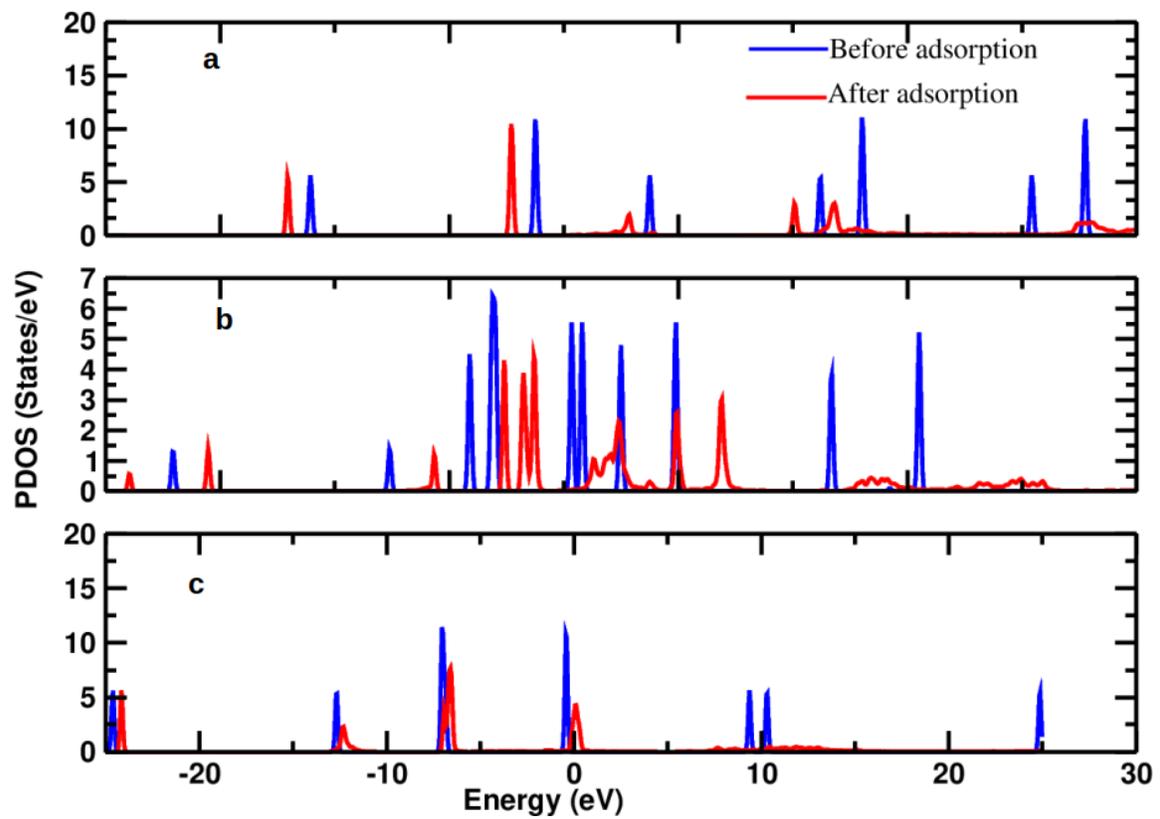

Figure 6: Calculated projected density of states of a) $NH_3$, b) $NO_2$ and c) NO on MoS/WTe heterostructure.

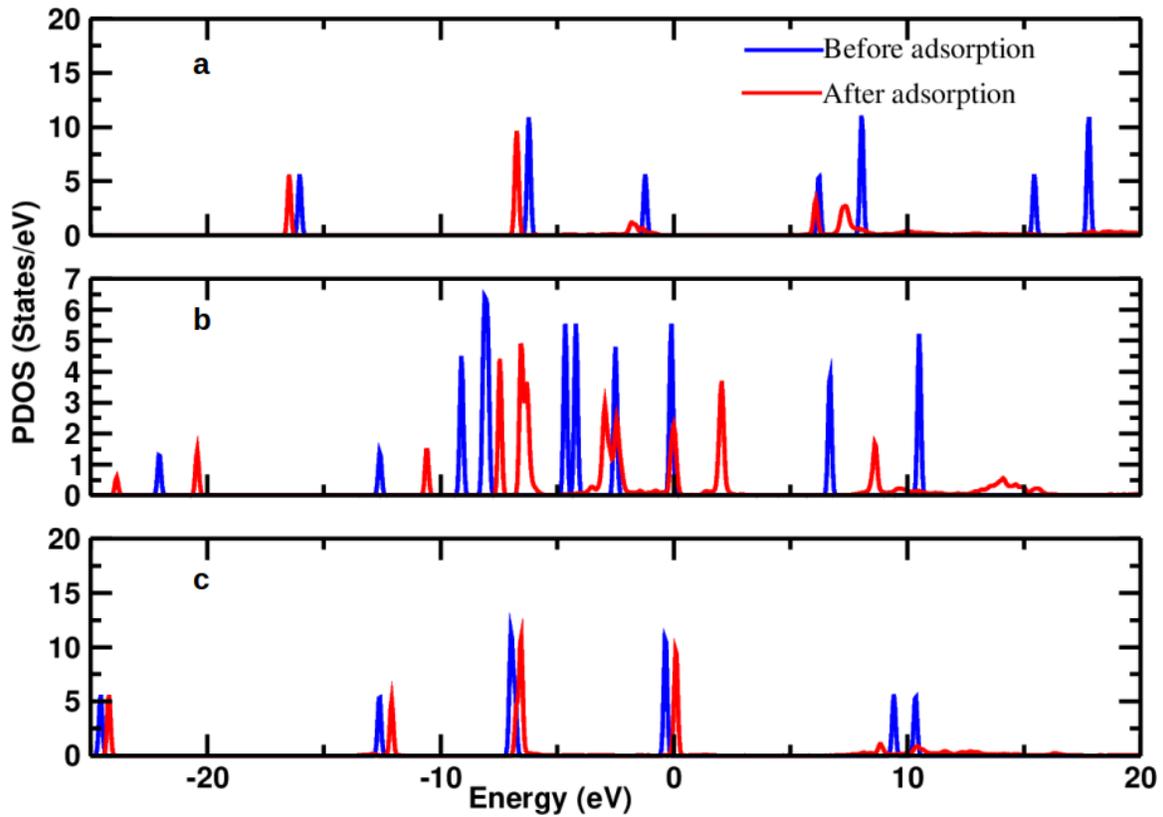

Figure 7: Calculated projected density of states of a) $NH_3$, b) $NO_2$ and c) NO on MoTe/WS heterostructure.

Thicker electron density accumulation is observed after the adsorption of NO on MoS/WTe and MoTe/WS heterostructures, whereas thicker electron density depletion is observed after the adsorption of NO on MoS/TiO and MoS/IrO heterostructures. The PDOS plot shows that small peaks form in the conduction band after NO adsorption on MoS/TiO and MoS/IrO heterostructures, indicating electron transfer to the heterostructures. This result is consistent with Bader charge calculations.

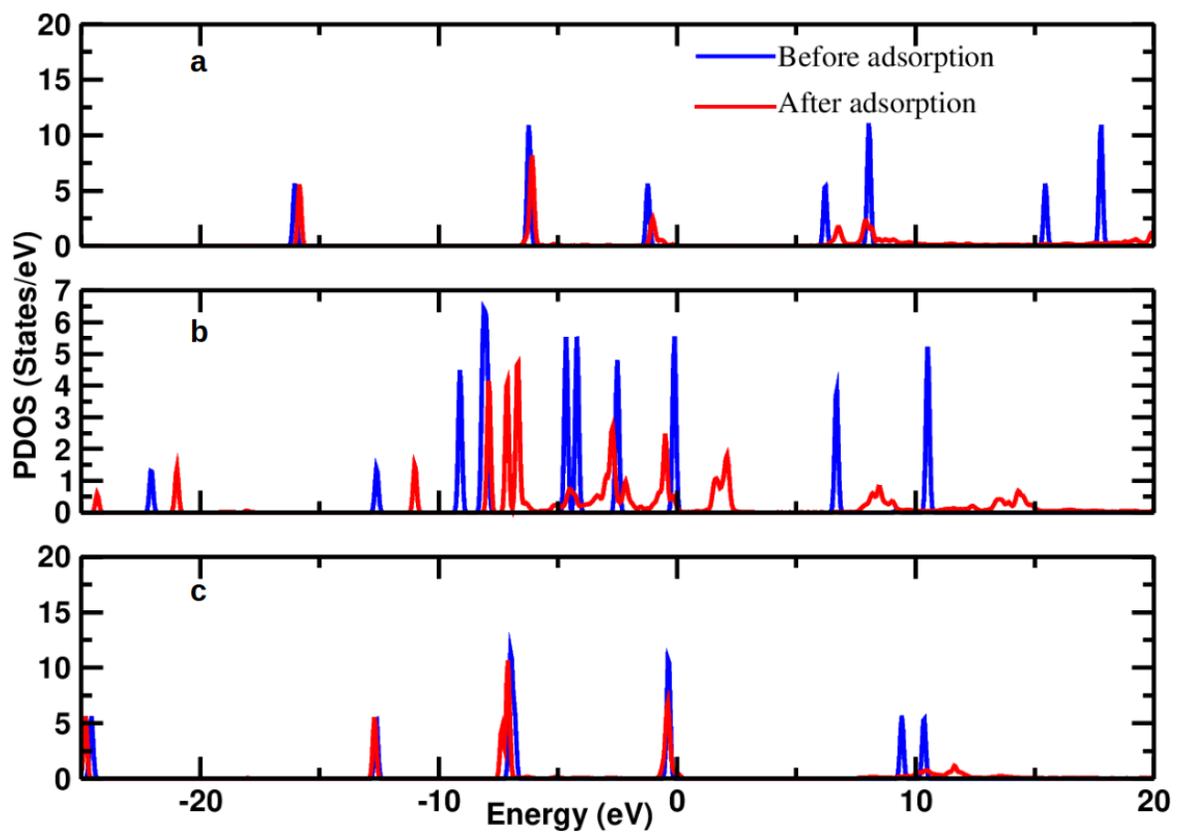

Figure 8: Calculated projected density of states of a) NH$_3$, b) NO$_2$ and c) NO on MoS/TiO heterostructure.

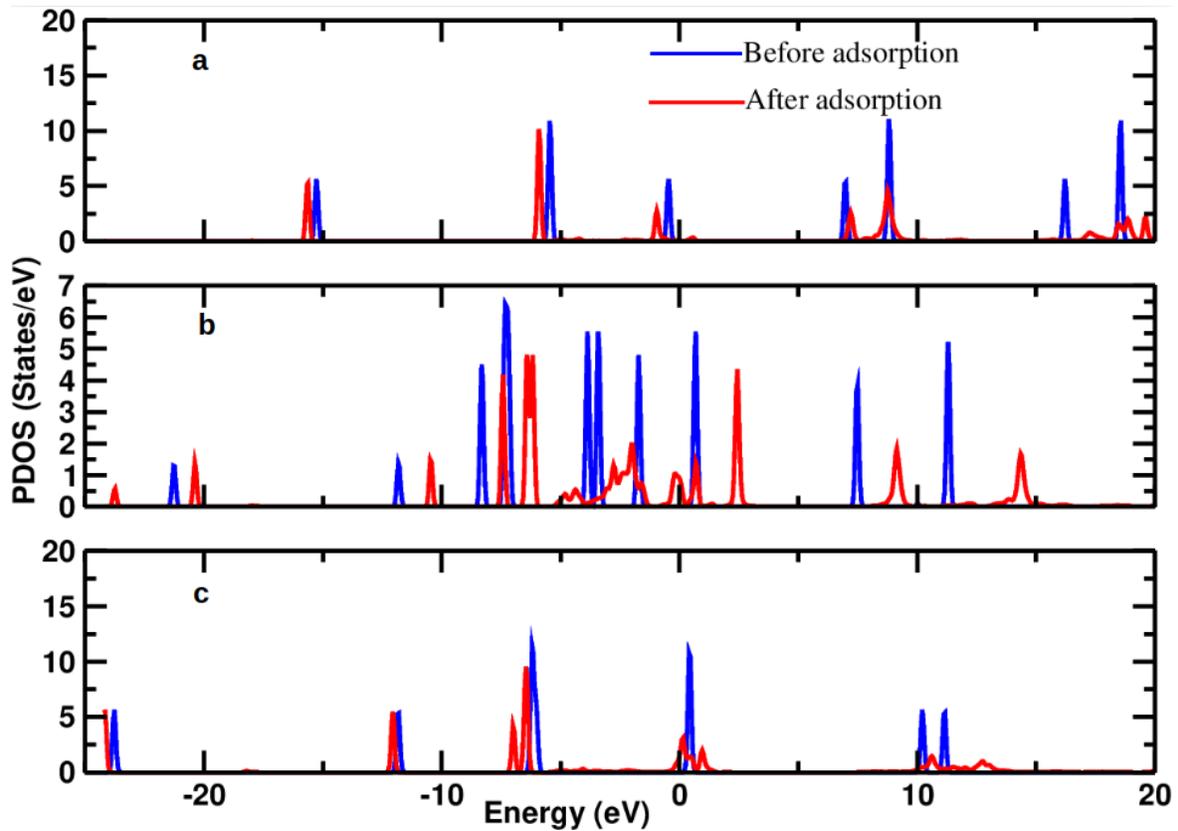

Figure 9: Calculated projected density of states of a) $NH_3$, b) $NO_2$ and c) NO on MoS/IrO heterostructure.

**Current (I)–Voltage (V) Characteristics**

To understand the sensing capabilities of the heterostructures for nitrogen- containing gas (NCG) molecules, their transmission properties were investigated. The transport properties of MoS/WTe, MoTe/WS, MoS/TiO, and MoS/IrO heterostructures, both with and without adsorbed NCGs, were studied using the TranSIESTA module implemented in the SIESTA code.

The transport setup was modeled using the same electrode materials (left and right), with both being semi-infinite. The scattering region, including NCGs molecules, was placed between them. Current was obtained using the Landauer-Büttiker formula in the bias voltage range (0.0, 1.0). Figure 10 displays the current-voltage characteristics of NCGs before and after adsorption on MoS/WTe heterostructure. The current responses are small both before and after adsorption of NCGs on MoS/WTe, MoTe/WS, MoS/TiO, and MoS/IrO heterostructures, and are within the same order of magnitude, except for the adsorption of $NH_3$ and NO on MoS/IrO heterostructure when applying a bias voltage of 1.0 V (Table S1). After the adsorption of $NH_3$, $NO_2$, and NO on MoS/WTe heterostructure, the current response increases with the applied voltage. However, after applying a bias of 0.8 V and 1.0 V, the current response remains constant for the adsorption of $NO_2$ on MoS/WTe heterostructure. Furthermore, the current response decreases after applying a bias voltage of 1.0 V for NO adsorption on MoS/WTe heterostructure. This result confirms that sensitivity starts at a low bias voltage of 0.1 V (indicating low power consumption), as most materials tend to exhibit sensitivity at higher voltages [52]. The current-voltage characteristic curves for MoTe/WS, MoS/IrO and MoS/TiO heterostructure with and without adsorbed gas molecules have been shown in Figure 10. The current responses show very small changes after adsorption of $NH_3$ and NO compared to the adsorption of $NO_2$ on MoTe/WS heterostructure.

Furthermore, the current response increases as the bias voltage increases, both with and without adsorption of NCGs on MoS/TiO heterostructure. However, the current response increases slightly after applying a bias voltage of 1.0 V in the adsorption of $NH_3$ and NO on MoS/TiO heterostructure. As shown in Table S1, the current response increases by one order of magnitude after applying a bias voltage of 1.0 V in the adsorption of $NH_3$ and NO on MoS/IrO heterostructure. The calculated sensitivities of $NH_3$ is lower compared to NO and $NO_2$ in MoS/WTe, MoTe/WS and MoS/TiO heterostructures, however, the calculated sensitivities of $NH_3$ and NO is higher than $NO_2$ adsorption in MoS/IrO heterostructure. As the voltage increases to 1.0 V the calculated sensitivities of $NH_3$ and NO increases one order of magnitude. Consequently, our results demonstrate that MoS/WTe, MoTe/WS and MoS/TiO heterostructures are more sensitive to the NOx molecules. The calculated sensitivities and selectivities of the gas sensing properties for nitrogen-containing gases on different heterostructures provide valuable information for their potential applications. The findings indicate that selectivity can be determined based on the calculated sensitivity of gas molecules. Specifically, in the MoS/WTe heterostructure, NO shows a higher sensitivity than $NH_3$ and $NO_2$ suggesting high selectivity for NO under these conditions. Additionally, $NO_2$ exhibits significantly high sensitivity in MoTe/WS and MoS/TiO heterostructures, while NO shows higher sensitivity at 1.0 V bias voltage. In the MoS/IrO heterostructure, NO displays higher sensitivity compared to $NH_3$ and $NO_2$. Overall, the results indicate that NOx has larger sensitivity in MoS/WTe, MoTe/WS, MoS/IrO, and MoS/TiO heterostructures at smaller applied bias voltage, leading to improved selectivity for NO and $NO_2$ molecules in these heterostructures. These findings have practical implications for the development of gas sensors with enhanced selectivity.

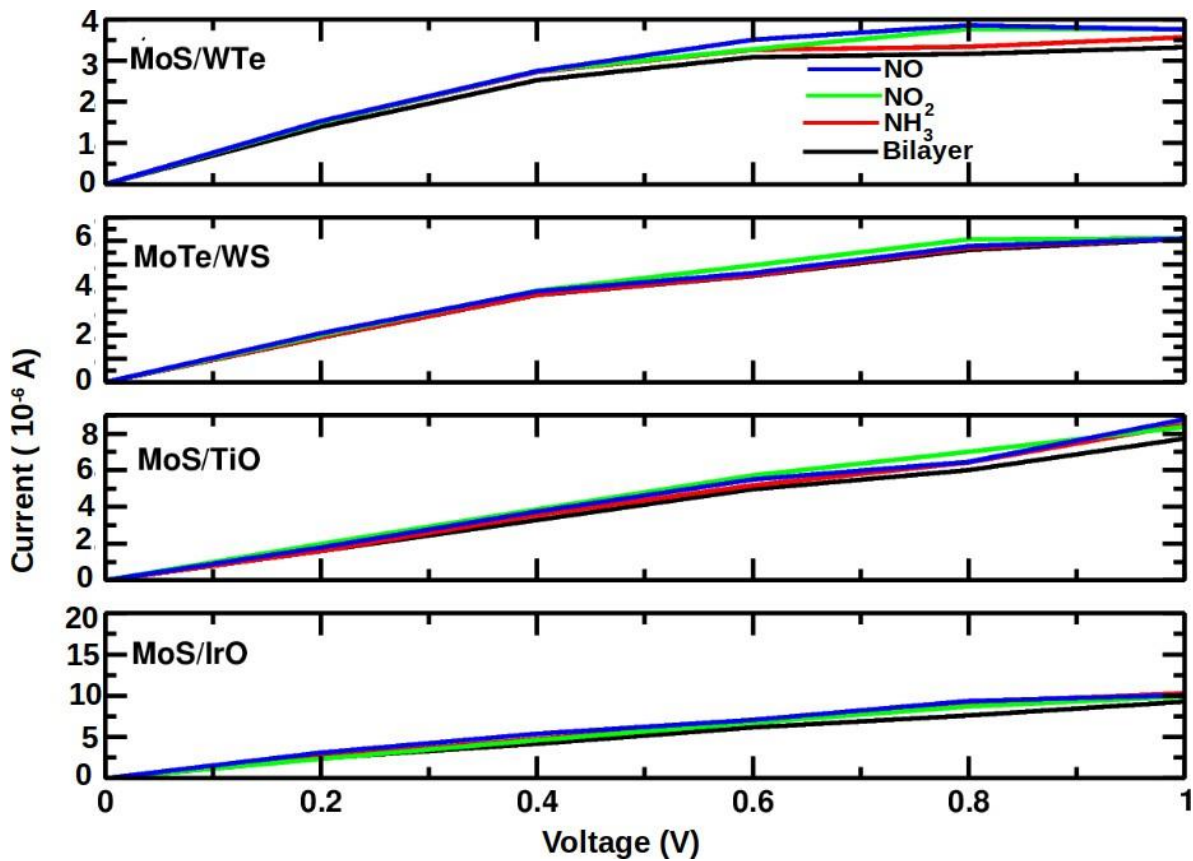

Figure 10: Current-Voltage curves for the adsorption of NCGs on bilayers (MoS/WTe, MoTe/WS, MoS/TiO and MoS/IrO)

**Band gap and Recovery time**

The relationship between band gap and electrical conductivity is an important factor in understanding the gas sensing properties of heterostructures. In our study, we investigated how the band gap influences the gas sensing properties of MoS/WTe, MoTe/WS, MoS/IrO, and MoS/TiO heterostructures for detecting nitrogen-containing gases (NCGs). The relationship

between band gap and electrical conductivity is described as follows [51, 52]:

$$\sigma \propto \exp\left(\frac{-E_g}{2kT}\right)$$

Where $\sigma$, k, Eg and T represents the electrical conductivity, the Boltzmann constant, the band gap and temperature, respectively.

The observed changes in band gap before and after the adsorption of nitrogen-containing gases (NCGs) on the various heterostructures (MoS/WTe and MoTe/WS) are crucial indicators of their sensitivity as gas sensors (Figure S1). These changes in band gap ranged from 0.1 eV to 0.35 eV in the adsorption of $NH_3$, $NO_2$, and NO. A significant change in band gap implies a corresponding alteration in the conductivity of the material. In this context, the observed band gap variations suggest that the conductivity of the heterostructures can be enhanced upon exposure to NCGs. This is a critical factor in determining the suitability of these materials for gas sensing applications. Evaluating the recovery time is a crucial aspect of assessing the performance of gas sensors. It provides valuable information about the sensor's ability to return to its initial state after being exposed to a specific gas. A shorter recovery time indicates a faster response and regeneration of the sensor, which is desirable for practical applications. The recovery time ($\tau$) can be calculated as follows:

$$\tau = v^{-1} \exp\left(\frac{-E_{ads}}{KT}\right)$$

Where k, T and $v^{-1}$, $E_{ads}$ is Boltzmann constant, temperature, the attempted frequency of the molecules and adsorption energy, respectively. It's valuable to know that the recovery times are shorter at room temperature in MoS/WTe and MoTe/WS heterostructures. Additionally, the observation of shorter recovery times for $NH_3$ in MoS/IrO and MoS/TiO heterostructures further highlights their potential for efficient desorption and sensor reusability. This means that $NH_3$ molecules have a tendency to desorb or detach from the surface of the heterostructures more quickly than $NO_2$ and NO molecules, indicating faster recovery times. It suggests that the gas sensors utilizing MoS/WTe and MoTe/WS bilayers have the potential to be efficiently recycled. As shown in Table S2, as the temperature increases the recovery times increases. This information is significant for practical applications, especially in gas sensing or catalytic processes. It's worth noting that these findings highlight the potential for these heterostructures in environmental and industrial applications where the detection and removal of $NH_3$, $NO_2$, and NO gases are important.

**Conclusion**

We conducted first-principle calculations to investigate the sensing properties of toxic gases, namely $NH_3$, $NO_2$, and NO, on MoS/WTe, MoTe/WS, MoS/IrO, and MoS/TiO heterostructures. Our findings indicate that $NH_3$ physisorbs onto MoS/WTe and MoTe/WS heterostructures, with adsorption energies of -0.68 eV and -0.54 eV for MoS/TiO and MoS/IrO heterostructures, respectively. The adsorption energies of $NO_2$ and NO are higher in MoS/IrO and MoS/TiO heterostructures. Bader charge analysis reveals that the charge transfer between the nitrogen-containing gases (NCGs) and the heterostructures is less than 0.5 electrons. The Bader charge analysis indicates that $NO_2$ and NO accept electrons in MoS/WTe and MoTe/WS heterostructures but lose electrons in MoS/TiO and MoS/IrO heterostructures. Electronic interactions were further investigated using Electron Density Difference (EDD) and Projected Density of States (PDOS) calculations between NCGs and the heterostructures. EDD and PDOS calculations also demonstrated substantial electron donation from MoS/WTe and MoTe/WS heterostructures to both NO and $NO_2$. Additionally, electron transport calculations showed small changes in current response before and after NCG adsorption on the heterostructures. The simulated current-voltage (I-V) curves of the NOx molecules on MoS/WTe, MoTe/WS, MoS/IrO, and MoS/TiO heterostructures showed that $NH_3$ induced only small changes in current response, while NO and $NO_2$ exhibited significant current response changes within a specific bias range. This suggests that the heterostructures have varying sensitivities to different gases, which is valuable information for potential gas sensing applications. Our study has identified specific heterostructures (MoS/WTe, MoTe/WS, MoS/TiO and MoS/IrO) with excellent potential for $NH_3$ detection, considering factors like adsorption energy, electronic property changes, and recovery times. In addition, MoS/WTe and MoTe/WS heterostructures can be a gas sensor for $NO_2$ and NO detection. This information is indeed valuable and could have significant implications in the development of sensitive and reliable gas sensors. The specificity and sensitivity of different heterostructures to specific toxic gases can have significant implications for real-world applications.

**CRediT authorship contribution statement**

JYD: Conceptualization, Data curation, Investigation, Methodology, Vi- sualization, Writing - orginal draft HA: Writing - review and editing

**Conflicts of interest**

There are no conflicts of interest to declare.

**Acknowledgment**

The research reported in this publication was supported by the Ministry of Education, Youth and Sports of the Czech Republic through the e- IN- FRA CZ ((ID:90254) and this work also supported by the project Quantum materials for applications in sustainable technologies (QM4ST), funded as project No. CZ.02.01.01/00/22_008/0004572 by Programme Johannes Amos Commenius, call Excellent Research.

**References**

[1] X. Tang, A. Du, L. Kou, Gas sensing and capturing based on two-dimensional layered materials: Overview from theoretical perspective, Wiley Interdisciplinary Reviews: Computational Molecular Science, 8 (2018) 1361.

[2] C.-H. Yeh, D.-W. Hsieh, Combined density functional theory calculation and non-equilibrium Green's function approach to predict the sensitivity of nitrogen-containing gases over PtTenS2-n monolayers (n= 0–2), FlatChem, 35 (2022) 100418.

[3] U. EPA, Integrated science assessment for oxides of nitrogen–health criteria, US Environmental Protection Agency, Washington, DC, 2016.

[4] T.J. Crowley, Causes of climate change over the past 1000 years, Science, 289 (2000) 270-277.

[5] W.H. Organization, World Health Statistics 2016 [OP]: Monitoring Health for the Sustainable Development Goals (SDGs), World Health Organization, 2016.

[6] W.H. Organization, A vision for primary health care in the 21st century: towards universal health coverage and the Sustainable Development Goals, in, World Health Organization, 2018.

[7] S. Kumar, V. Pavelyev, P. Mishra, N. Tripathi, P. Sharma, F. Calle, A review on 2D transition metal di-chalcogenides and metal oxide nanostructures based $NO_2$ gas sensors, Materials Science in Semiconductor Processing, 107 (2020) 104865.

[8] S.M. Bernard, J.M. Samet, A. Grambsch, K.L. Ebi, I. Romieu, the potential impacts of climate variability and change on air pollution-related health effects in the United States,

Environmental health perspectives, 109 (2001) 199-209.

[9] Y. Chen, L. Zhang, Y. Zhao, L. Zhang, J. Zhang, M. Liu, M. Zhou, B. Luo, High-resolution ammonia emissions from nitrogen fertilizer application in China during 2005–2020, Atmosphere, 13 (2022) 1297.

[10] S.S. Dindorkar, A. Yadav, Monolayered silicon carbide for sensing toxic gases: a comprehensive study based on the first-principle density functional theory, Silicon, 14 (2022) 11771-11779.

[11] T.W. Hesterberg, W.B. Bunn, R.O. McClellan, A.K. Hamade, C.M. Long, P.A. Valberg, Critical review of the human data on short-term nitrogen dioxide (NO2) exposures: evidence for NO2 no-effect levels, Critical reviews in toxicology, 39 (2009) 743-781.

[12] E. Salih, A.I. Ayesh, First principle study of transition metals codoped $MoS_2$ as a gas sensor for the detection of NO and $NO_2$ gases, Physica E: Low-dimensional Systems and Nanostructures, 131 (2021) 114736.

[13] J. Warner, R. Dickerson, Z. Wei, L.L. Strow, Y. Wang, Q. Liang, increased atmospheric ammonia over the world's major agricultural areas detected from space, Geophysical Research Letters, 44 (2017) 2875-2884.

[14] Y.Y. Broza, R. Vishinkin, O. Barash, M.K. Nakhleh, H. Haick, Synergy between nanomaterials and volatile organic compounds for non-invasive medical evaluation, Chemical Society Reviews, 47 (2018) 4781-4859.

[15] V. Galstyan, M.P. Bhandari, V. Sberveglieri, G. Sberveglieri, E. Comini, Metal oxide nanostructures in food applications: Quality control and packaging, Chemosensors, 6 (2018) 16.

[16] A.H. Jalal, F. Alam, S. Roychoudhury, Y. Umasankar, N. Pala, S. Bhansali, Prospects and challenges of volatile organic compound sensors in human healthcare, Acs Sensors, 3 (2018) 1246-1263.

[17] L.A. Mercante, R.S. Andre, L.H. Mattoso, D.S. Correa, Electrospun ceramic nanofibers and hybrid-nanofiber composites for gas sensing, ACS Applied Nano Materials, 2 (2019) 4026-4042.

[18] T.R. Pavase, H. Lin, S. Hussain, Z. Li, I. Ahmed, L. Lv, L. Sun, S.B.H. Shah, M.T. Kalhoro, Recent advances of conjugated polymer (CP) nanocomposite-based chemical sensors and their applications in food spoilage detection: A comprehensive review, Sensors and Actuators B: Chemical, 273 (2018) 1113-1138.

[19] V. Babar, H. Vovusha, U. Schwingenschlögl, Density functional theory analysis of gas adsorption on monolayer and few layer transition metal dichalcogenides: Implications for sensing, ACS Applied Nano Materials, 2 (2019) 6076-6080.

[20] H. Cui, G. Zhang, X. Zhang, J. Tang, Rh-doped $MoSe_2$ as a toxic gas scavenger: A first-principles study, Nanoscale Advances, 1 (2019) 772-780.

[21] N. Goel, K. Kunal, A. Kushwaha, M. Kumar, Metal oxide semiconductors for gas sensing, Engineering Reports, 5 (2023) 12604.

[22] H. Ji, W. Zeng, Y. Li, Gas sensing mechanisms of metal oxide semiconductors: a focus review, Nanoscale, 11 (2019) 22664-22684.

[23] Y.-S. Shim, K.C. Kwon, J.M. Suh, K.S. Choi, Y.G. Song, W. Sohn, S. Choi, K. Hong, J.-M. Jeon, S.-P. Hong, Synthesis of numerous edge sites in $MoS_2$ via $SiO_2$ nanorods platform for highly sensitive gas sensor, ACS applied materials & interfaces, 10 (2018) 31594-31602.

[24] M.J. Szary, D.M. Florjan, J.A. Bąbelek, Selective detection of carbon monoxide on P-block doped monolayers of MoTe$_2$, ACS sensors, 7 (2022) 272-285.

[25] Z. Xu, Adsorption and sensing mechanisms of Ni-doped PtTe$_2$ monolayer upon NO$_2$ and O$_3$ in air-insulated switchgears, RSC advances, 13 (2023) 18129-18137.

[26] S.-Y. Cho, S.J. Kim, Y. Lee, J.-S. Kim, W.-B. Jung, H.-W. Yoo, J. Kim, H.-T. Jung, Highly enhanced gas adsorption properties in vertically aligned MoS2 layers, ACS nano, 9 (2015) 9314-9321.

[27] Y. Kim, S.-K. Kang, N.-C. Oh, H.-D. Lee, S.-M. Lee, J. Park, H. Kim, Improved sensitivity in Schottky contacted two-dimensional MoS2 gas sensor, ACS applied materials & interfaces, 11 (2019) 38902-38909.

[28] Z. Meng, R.M. Stolz, L. Mendecki, K.A. Mirica, Electrically-transduced chemical sensors based on two-dimensional nanomaterials, Chemical reviews, 119 (2019) 478-598.

[29] N. Joshi, M.L. Braunger, F.M. Shimizu, A. Riul, O.N. Oliveira, Two-dimensional transition metal dichalcogenides for gas sensing applications, Nanosensors for environmental applications, (2020) 131-155.

[30] Y. Kim, S. Lee, J.G. Song, K.Y. Ko, W.J. Woo, S.W. Lee, M. Park, H. Lee, Z. Lee, H. Choi, 2D transition metal dichalcogenide heterostructures for p-and n-type photovoltaic self-powered gas sensor, Advanced Functional Materials, 30 (2020) 2003360.

[31] E. Lee, Y.S. Yoon, D.-J. Kim, Two-dimensional transition metal dichalcogenides and metal oxide hybrids for gas sensing, ACS sensors, 3 (2018) 2045-2060.

[32] B.L. Li, J. Wang, H.L. Zou, S. Garaj, C.T. Lim, J. Xie, N.B. Li, D.T. Leong, Low-dimensional transition metal dichalcogenide nanostructures-based sensors, Advanced Functional Materials, 26 (2016) 7034-7056.

[33] J. Ping, Z. Fan, M. Sindoro, Y. Ying, H. Zhang, Recent advances in sensing applications of two-dimensional transition metal dichalcogenide nanosheets and their composites, Advanced Functional Materials, 27 (2017) 1605817.

[34] D.J. Late, T. Doneux, M. Bougouma, Single-layer $MoSe_2$ based $NH_3$ gas sensor, Applied physics letters, 105 (2014).

[35] P. Panigrahi, T. Hussain, A. Karton, R. Ahuja, Elemental substitution of two-dimensional transition metal dichalcogenides ($MoSe_2$ and $MoTe_2$): implications for enhanced gas sensing, ACS sensors, 4 (2019) 2646-2653.

[36] W. Yuan, A. Liu, L. Huang, C. Li, G. Shi, High-performance NO2 sensors based on chemically modified graphene, Advanced Materials (Deerfield Beach, Fla.), 25 (2012) 766-771.

[37] J. Baek, D. Yin, N. Liu, I. Omkaram, C. Jung, H. Im, S. Hong, S.M. Kim, Y.K. Hong, J. Hur, A highly sensitive chemical gas detecting transistor based on highly crystalline CVD-grown $MoSe_2$ films, Nano Research, 10 (2017) 1861-1871.

[38] B. Cho, M.G. Hahm, M. Choi, J. Yoon, A.R. Kim, Y.-J. Lee, S.-G. Park, J.-D. Kwon, C.S. Kim, M. Song, Charge-transfer-based gas sensing using atomic-layer $MoS_2$, Scientific reports, 5 (2015) 8052.

[39] F.K. Perkins, A.L. Friedman, E. Cobas, P. Campbell, G. Jernigan, B.T. Jonker, Chemical vapor sensing with monolayer $MoS_2$, Nano letters, 13 (2013) 668-673.

[40] F. Schedin, A.K. Geim, S.V. Morozov, E.W. Hill, P. Blake, M.I. Katsnelson, K.S. Novoselov, Detection of individual gas molecules adsorbed on graphene, Nature materials, 6 (2007) 652-655.

[41] F. Urban, F. Giubileo, A. Grillo, L. Iemmo, G. Luongo, M. Passacantando, T. Foller,

L. Madauß, E. Pollmann, M.P. Geller, Gas dependent hysteresis in $MoS_2$ field effect transistors, 2D Materials, 6 (2019) 045049.

[42] J.Y. Damte, J. Houska, Tribo-piezoelectric nanogenerators for energy harvesting: a first-principles study, Sustainable Energy & Fuels, 2024.

[43] E.A. Jose MSoler, Julian D Gale, Alberto Garcia, Javier Junquera, Pablo Ordejon, Daniel Sanchez, The SIESTA method for ab initio order-N materials simulation, J. Phys.: Condens. Matter, 14 (2002) 2745–2779.

[44] J. Klimeš, D.R. Bowler, A. Michaelides, Van der Waals density functionals applied to solids, Physical Review B—Condensed Matter and Materials Physics, 83 (2011) 195131.

[45] J.K.A. Obligacion, D.B. Putungan, 2D 1T′-$MoS_2$-$WS_2$ van der Waals heterostructure for hydrogen evolution reaction: dispersion-corrected density functional theory calculations, Materials Research Express, 7 (2020) 075506.

[46] N. Papior, N. Lorente, T. Frederiksen, A. García, M. Brandbyge, Improvements on non-equilibrium and transport Green function techniques: The next-generation transiesta, Computer Physics Communications, 212 (2017) 8-24.

[47] O.B. Malcıoglu, Ş. Erkoç, Functionality of C (4, 4) carbon nanotube as molecular detector, Journal of nanoscience and nanotechnology, 8 (2008) 469-478.

[48] M. Büttiker, Y. Imry, R. Landauer, S. Pinhas, Generalized many-channel conductance formula with application to small rings, Physical Review B, 31 (1985) 6207.

[49] D. Chakraborty, P. Johari, First-principles investigation of the 1T-$HfTe_2$ nanosheet for selective gas sensing, ACS Applied Nano Materials, 3 (2020) 5160-5171.

[50] Q. Yue, Z. Shao, S. Chang, J. Li, Adsorption of gas molecules on monolayer $MoS_2$

and effect of applied electric field, Nanoscale research letters, 8 (2013) 1-7.

[51] A. Ahmadi, N.L. Hadipour, M. Kamfiroozi, Z. Bagheri, Theoretical study of aluminum nitride nanotubes for chemical sensing of formaldehyde, Sensors and Actuators B: Chemical, 161 (2012) 1025-1029.

[52] C.-H. Yeh, W.-Y. Lin, J.-C. Jiang, Enhancement of chlorobenzene sensing by doping aluminum on nanotubes: A DFT study, Applied Surface Science, 514 (2020) 145897.